\documentclass[conference]{IEEEtran}
\usepackage{graphicx}
\usepackage{multirow}
\usepackage{amsmath}
\usepackage{algorithmic}
\usepackage{algorithm}

\newtheorem{theorem}{Theorem}

\newtheorem{definition}{Definition}[section]

\usepackage{amssymb}
\usepackage{multirow}

\newtheorem{example}{Example}[section]

\newtheorem{proposition}{Prop.}[section]

\newcommand{\nop}[1]{{}}

\def\calm{{\mathcal{M}}}
\def\Act{\mathcal{A}}
\def\APPall{\Act\times\calm}

\def\APP{\Phi}

\def\BV{\textsf{B}}
\def\CM{\textsf{C}}
\def\bc{\textbf{\textsf{c}}}

\def\setS{\textbf{S}}

\def\calg{\mathcal{G}}

\def\calh{\mathcal{H}}

\def\call{\mathcal{L}}

\def\exm{\calm_{cpgn}}
\def\exs{s_{cpgn}}
\def\exb{\BV_{cpgn}}
\def\exA{\Act_{cpgn}}

\def\exc{\CM_{cpgn}^{(s)}}
\def\excs{\CM_{cpgn}^{(\exs)}}
\def\exic{IC_{cpgn}}

\def\inset{\Theta_{in}}
\def\outset{\Theta_{out}}
\def\insetex{\Theta_{in}^{cpgn}}

\def\appl{appl}
\def\Qset{Q^{\Gamma}_{(a,p)}}
\def\Qseti{Q^{\Gamma}_{(a_i,p_i)}}
\def\Qsetj{Q^{\Gamma}_{(a_j,p_j)}}
\newtheorem{lemma}{Lemma}[section]
\def\Affset{\textsf{Aff}_{(a,p)}^{\Gamma}}
\def\Affseti{\textsf{Aff}_{(a_i,p_i)}^{\Gamma}}
\def\Affsetj{\textsf{Aff}_{(a_j,p_j)}^{\Gamma}}

\begin{document}

\title{Geospatial Optimization Problems}
\author{\IEEEauthorblockN{Paulo Shakarian}
\IEEEauthorblockA{Network Science Center\\
Dept. of Electrical Engineering and\\
Computer Science\\
U.S. Military Academy\\
West Point, NY 10996\\
paulo[at]shakarian.net}
\and
\IEEEauthorblockN{V.S. Subrahmanian}
\IEEEauthorblockA{Dept. of Computer Science\\
University of Maryland\\
College Park MD \\
vs[at]cs.umd.edu}}



\maketitle

\begin{abstract}
\noindent There are numerous applications which require the ability to take certain actions (e.g. distribute money, medicines, people etc.) over a geographic region. A disaster relief organization must allocate people and supplies to parts of a region after a disaster. A public health organization must allocate limited vaccine to people across a region. In both cases, the organization is trying to optimize something (e.g. minimize expected number of people with a disease). We introduce  ``geospatial optimization problems'' (GOPs) where an organization has limited resources and budget to take actions in a geographic area.  The actions result in one or more properties changing for one or more locations. There are also certain constraints on the combinations of actions that can be taken. We study two types of GOPs - goal-based and benefit-maximizing (GBGOP and BMGOP respectively).  A GBGOP ensures that certain properties must be true at specified locations after the actions are taken while a BMGOP optimizes a linear benefit function. We show both problems to be NP-hard (with membership in NP for the associated decision problems).  Additionally, we prove limits on approximation for both problems.  We present integer programs for both GOPs that provide exact solutions.  We also correctly reduce the number of variables in for the GBGOP integer constraints.  For BMGOP, we present the \textsf{BMGOP-Compute} algorithm that runs in PTIME and provides a reasonable approximation guarantee in most cases. 
\end{abstract}

\section{Introduction}
\label{introduction}
As geo-located social network data becomes more common with sites such as FourSquare\footnote{https://foursquare.com/} and programs such as RealityMining\footnote{http://realitycommons.media.mit.edu/}, it becomes desirable to reason about such data.  There are numerous applications which require the ability to take certain actions (e.g. distribute money, medicines, people etc.) over a geographic region. For instance, a disaster relief organization must allocate people and supplies in a region after a disaster.  A public health organization needs to allocate limited vaccine stocks to people across the region.
A government needs to allocate funds for education or unemployment training across a region.  However, allocating any resource will cause certain effects - some desirable, some not - based on the network connections among geographic locations.  In this paper we present a formal framework that allows reasoning about such geo-located data in order to answer certain queries where we have some desired goal to achieve as the result of our geographically-based resource allocation - \textbf{all the while considering the complex interactions among locations.}

\begin{figure}
	\centering
		\includegraphics[scale=0.5]{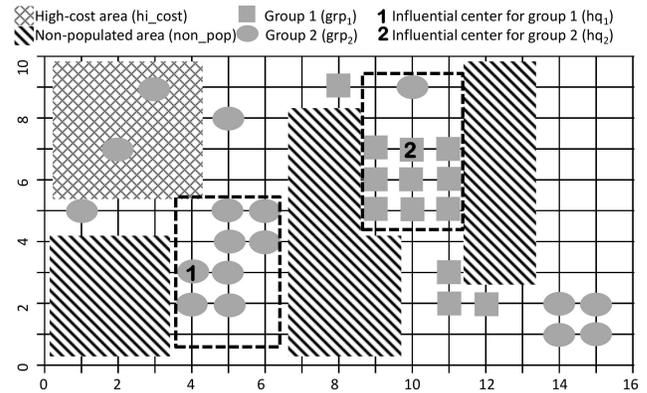}
	\caption{Locations in a district - contingency groups and unpopulated areas.}
	\label{fig1}
\end{figure}

Figure~\ref{fig1} shows a 2-dimensional map of a region.  A political candidate can only make so many campaign stops and public appeals.  We assume that a map $\calm$  is discrete (this is a common assumption  in most GIS systems) and has coordinates drawn from $[0,\ldots,M]\times [0,\ldots N]$ where the bottom left corner of the map is the point $(0,0)$.  The candidate wants to identify the best places to campaign or make public appeals to maximize his exposure.  Additionally, the map shows un-populated areas, areas where campaigning costs are high, and areas dominated by one of two constituent groups.  All of these factors may affect the set of locations the candidate selects to optimize his exposure.

In this paper, we introduce \emph{geographic optimization problems} or GOPs that capture and solve problems such as those mentioned above.  This framework allows one to more prudently position resources in a manner to achieve a goal while considering the complex interactions between locations (that may be modeled as a network).  The organization and contribution of the paper is as follows.  Section~\ref{tech-prelims} formally defines GOPs - specifically we introduce goal-based and benefit-maximizing GOPs (GBGOP and BMGOP respectively). Section~\ref{complexity:sec} shows that both GBGOP and BMGOP are NP-hard (with the associated decision problems in the complexity class NP).  Additionally, we prove non-trivial theoretical limits on approximation: if GBGOP were to be approximated within the logarithm of the input then NP would have a slightly super-polynomial oracle.  BMGOP cannot be approximated within a guaranteed factor greater than 0.63 unless P=NP.  Section~\ref{algo:sec} presents integer programs to solve both GBGOP and BMGOP using an IP solver like CPLEX.   In Section~\ref{var-red-sec}, we show how to correctly reduce the number of variables in the integer constraints for GBGOP.  We then develop the \textsf{BMGOP-Compute} algorithm in Section~\ref{mu-alg-sec} that can quickly approximate a BMGOP in polynomial time and provides an approximation guarantee.

\section{GOPs Formalized}
\label{tech-prelims}
Throughout this paper, we assume that $\calm=[0,\ldots,M]\times[0,\ldots,N]$ is an arbitrary, but fixed ``map''.
We define a logical language $\call$ whose constant symbols are members of $\calm$ and that has an infinite set $\call_{var}$ of variable symbols disjoint from $\calm$.  $\call$ has a set $\calg = \{ g_1,\ldots,g_n\}$ of  unary predicate symbols.  As usual, a term is either a constant symbol or variable symbol. If $t$ is a term, then $g_i(t)$ is an \emph{atom}. If $t$ is a constant, then $g_i(t)$ is \emph{ground}. Intuitively, if $p\in\calm$, then $g_i(p)$ says that point $p$ has property $g_i$.   We use $B_\call$ to denote the set of all ground atoms.  Well-formed formulas (wffs) are defined in the usual way. (i) Every atom is a wff. (ii) If $F,G$ are wffs, then so are $F\,\wedge\, G, F\,\lor\, G,\neg F$ are all wffs.

\begin{example}
\label{first-ex}
Consider the map $\exm$ in Figure~\ref{fig1} with predicates $\calg=\{ hi\_cost, non\_pop, grp_1, grp_2, hq_1, hq_2\}$.  The predicate $exposure$ not depicted in the figure corresponds to a candidate receiving exposure in a certain area.   $hi\_cost((1,9)), hq_1((4,3)), non\_pop((8,1)),$ and $grp_2((5,8))$ are all examples of ground atoms.
\end{example}

A \emph{state} is any subset of $B_\call$.   We use $\setS$ to denote the set of all states.  Satisfaction of formulas is defined in the obvious way. State $s$ satisfies a ground atom $A$, denoted $s\models A$,  iff $A\in s$. $s\models F\,\lor\, G$ iff $s\models F$ or $s\models G$. $s\models F\,\wedge\, G$ iff $s\models F$ and $s\models G$. $s\models \neg F$ iff $s$ does not satisfy $F$.

\begin{example}
\label{st-ex}
The shading shown in Figure~\ref{fig1} defines a state.  For example, $hi\_cost((1,9)) \in \exs$ while $exposure((1,9))\notin \exs$.
\end{example}

An action maps points to sets of ground atoms. 

\begin{definition}[Action]
An action is a mapping $a: \calm \rightarrow 2^{B_\call}$. We use $\Act$ to denote the set of actions.  An action-point pair is any member of $\APPall$.
\end{definition}
An action-point pair $(a,p)$ is executed if action $a$ takes place at point $p$. Thus, one can think of $(a,p)$ as saying that action $a$ occurs at point $p$.  The \emph{result of executing a set $SOL$ of action-point pairs in state $s_0$ is denoted
$appl(SOL,s_0)$ and is the set 
$(s_0\,\cup\,\{ a(p)\: |\: (a,p)\in SOL\})$.}

\begin{example}
Continuing with example~\ref{bf-ex}, our candidate has actions $\exA=\{nor, appeal_1, appeal_2 \}$ where $nor$ refers to a normal campaign stop and $appeal_1, appeal_2$ refer to public appeals to constituent groups 1 and 2 respectively.  The actions map to ground atoms as follows.
\begin{small}
\begin{eqnarray*}
nor(p) = & \{exposure(p') | & \neg non\_pop(p') \wedge d(p,p') \leq 1 \}\\
appeal_i(p) = & \{exposure(p') | & hq_i(p) \wedge grp_i(p')\}
\end{eqnarray*}
\end{small}
The first action says that when a normal campign stop is made at point $p$ and $p'$ is a populated place one distance unit or less from $p$, then the candidate has exposure at place $p'$ as well. The second action says that if the candidate makes an appeal (action) at point $p$ and $p$ is the headquarters of interest group $grp_i$, then the candidate has obtained exposure in all places associated with interest group $grp_i$.
\end{example}


\begin{definition}[Cost Function]
A \textbf{cost function}, $\CM : \APPall \rightarrow [0,1]$.
\end{definition}
Throughout this paper, we assume the cost function is arbitrary but fixed and can be computed in constant time.   We also assume that if $\APPall = \{ (a_1,p_1),\ldots,(a_m,p_m)\}$, then $c_i$ is used to denote $\CM(a_i,p_i)$.

\begin{example}
\label{cost-ex}
The cost function for our example is $\exc$ and is defined (based on some state $s$) as follows: $\exc(a,p)=1$ if $hi\_cost(p) \in s$ and $0.5$ otherwise.
\end{example}

We also assume the existence of a set of integrity constraints $IC$ that specify that certain actions cannot be jointly taken if some conditions hold w.r.t. the state --- such constraints were defined before by
\cite{eiter99}.

\begin{definition}[Integrity Constraint]
If $\APP$ is a set of action-point pairs and $\chi$ is a wff, then $\APP \hookleftarrow \chi$ is an \textbf{integrity constraint}.
\end{definition}
When $\APP \hookleftarrow \chi$  is ground, this says that if $\chi$ is true, then only one action-point pair in $\APP$ may be executed. Formally, suppose $s$ is a state and $\APP'$ is a set of action-point pairs and $\APP \hookleftarrow \chi$  is ground. $(s, \APP') \models \APP \hookleftarrow \chi$ iff either $s \not\models \chi$ or $s \models \chi$ \textbf{and} $| \APP \cap \APP' | \leq 1$.   $(s,\APP')$ satisfies an integrity constraint iff it satisfies all ground instances of it. $(s, \APP') \models IC$ where $IC$ is a set of  integrity constraints iff $(s, \APP')$ satisfies every constraint in that set.  Given a state $s$ and set $IC$ of integrity constraints, we use $IC_{s}$ to denote the set of all ground instances of integrity constraints in $IC$ where the associated wff $\chi$ is satisfied by $s$\footnote{Formally, $IC_{s} = \{(\APP \hookleftarrow \chi) \in IC | s \models \chi \}$}.

\begin{example}
\label{ic-ex}
Continuing Example~\ref{cost-ex}, let $\exic$ be $\{\{appeal_1((4,3)),appeal_2((10,7))\}\hookleftarrow \textsf{TRUE}\}$.  This constraint says that an appeal can be made to either group 1 or group 2 at their center of influence, but not both --- for instance, these two groups may have opposing views.
\end{example}

We now introduce the \textit{goal-based geospatial optimization problem} (GBGOP).  This problem takes as input a map $\calm$, initial state $s_0$, set of actions $\Act$, cost function $\CM$, integrity constraints $IC$, positive real number $\bc$, and disjoint sets $\inset,\outset \subseteq B_\call$.  Intuitively, $\bc$ restricts the total cost and $\inset$ (resp. $\outset$) is a set of atoms that must be true (resp. false) after the actions are applied.  Our optimality criteria for a GBGOP is to minimize the cardinality of the action-point pairs.  A GBGOP can be viewed as an abductive inference problem (i.e. find a set of actions that lead to the current state) - where minimal cardinality is a common parsimony requirement.

\begin{definition}[GBGOP Solution, Optimal Solution]
A \emph{solution} to a GBGOP $(\calm,s_0,\Act,\CM,IC,\bc,\inset,\outset)$ is a set $SOL\subseteq  \APPall$  such that:
\textsf{(i)} $\Sigma_{(a_i,p_i)\in SOL} c_{i} \leq \bc$, \textsf{(ii)} $(s_0, SOL) \models IC$, and \textsf{(iii)} $\appl(s_0, SOL) \models \bigwedge_{A_i \in \inset}A_i \wedge \bigwedge_{A_j \in \outset}\neg A_j$. 

A solution $SOL$ is \emph{optimal} iff there is no other solution $SOL'$ such that $|SOL'| \leq |SOL|$.
\end{definition}

Our next type of problem is a \emph{benefit-maximizing geospatial optimization problem} (BMGOP) that also considers a benefit function, defined as follows.

\begin{definition}[Benefit Function]
The \textbf{benefit function}, $\BV : B_\call \rightarrow \Re^{+}$ maps atoms to positive real numbers.  
\end{definition}

\begin{example}
\label{bf-ex}
In our running example, we use the benefit function $\exb$ where $\exb(A) = 1$ if $A$ has the form $exposure()$ and $0$ otherwise.
\end{example}

As with cost, we assume the benefit function to be arbitrary but fixed and computable in constant time.   We also assume that if $B_\call=\{ A_1,\ldots,A_n\}$, then $\BV(A_i)$ is denoted $b_i$.  A BMGOP takes as input, $\calm$, $s_0$, $\Act$, $\CM$, $IC$, and $\bc$ - all defined the same as for a GBGOP.  Additionally it takes benefit function $\BV$ and natural number $k$.  Here $k$ is a bound on the number of actions the agent can take as we attempt to maximize benefit as an optimality criteria.

\begin{definition}[BMGOP Solution, Optimal Solution]
A \emph{solution} to a BMGOP $(\calm,s_0,\BV,\Act,\CM,IC,k,\bc)$ is a set $SOL\subseteq  \APPall$  such that:
\textsf{(i)} $|SOL|\leq k$ and \textsf{(ii)} $\Sigma_{(a_i,p_i)\in SOL} c_{i} \leq \bc$, and \textsf{(iii)}
$(s_0, SOL) \models IC$. 

A solution $SOL$ is \emph{optimal} iff there is no other solution $SOL'$ such that 
$\sum_{A_i \in appl(SOL,s_0)}b_i < \sum_{A_i \in appl(SOL',s_0)}b_i$.
\end{definition}
\section{Complexity Results}\label{complexity:sec}
Here, we provide complexity results for GBGOPs and BMGOPs. First, we establish both as being at least NP-hard.

\begin{theorem}
\label{gbgop-nph}
Given GBGOP $(\calm,s_0,\Act,\CM,IC,\bc,\inset,$\\$\outset)$, finding an optimal solution $SOL \subseteq \APPall$ is NP-hard.
This result holds even if for each $a \in \Act, p \in \calm$, it is the case that $\forall g'(p') \in a(p)$, $p' = p$ - i.e. each action only affects the point is is applied to.
\end{theorem}
\noindent\textit{Proof Sketch.} We embed the known NP-hard problem of SET-COVER~\cite{feige98} which takes as input a set of $n$ elements, $S$ and a family of $m$ subsets of $S$, $\calh \equiv \{ H_1, \ldots, H_{m} \}$, and outputs $\calh' \subseteq \calh$ s.t. the union of the subsets covers all elements in $S$ and $\calh'$ is of minimal cardinality.  We encode this problem into a GBGOP as follows: we set $\calg = \{g_1,\ldots,g_n\}$ - each predicate in $\calg$ corresponds to an element in $S$, the map, $\calm$ consists of a single point, $p$, the actions $\Act = \{a_1,\ldots,a_m\}$ s..t each action $a_i\Act$ corresponds to an element in $\calh$ and each is defined as follows: $a_i(p) = \bigcup_{x_j \in H_i}\{g_j(p) \}$.  The cost function $\CM$ returns $1$ for each action-point pair, $\inset = \bigcup_{g_i \in \calg}\{ g_i(p)\}$, $\outset = \emptyset$, and finally, we set $s_0 = \emptyset$, $IC = \emptyset$, $\bc = n$.$\hfill\Box$\\

\begin{theorem}
\label{gop-nph}
Given BMGOP $(\calm,s_0,\BV,\Act,\CM,IC,k,\bc)$, finding an optimal solution $SOL \subseteq \Act$ is NP-hard.
This result holds even if for each $a \in \Act, p \in \calm$, it is the case that $\forall g'(p') \in a(p)$, $p' = p$ - i.e. each action only affects the point is is applied to).
\end{theorem}
\noindent\textit{Proof Sketch.} We embed the known NP-hard problem of MAX-K-COVER~\cite{feige98} which takes as input a set of $n$ elements, $S$ and a family of $m$ subsets of $S$, $\calh \equiv \{ H_1, \ldots, H_{m} \}$, and positive integer $K$ and outputs $\leq K$ subsets from $\calh$ s.t. the union of the subsets covers a maximal number of elements in $S$.  We encode this problem into a BMGOP as follows: we set $\calg = \{g_1,\ldots,g_n\}$ - each predicate in $\calg$ corresponds to an element in $S$, the map, $\calm$ consists of a single point, $p$, the function $\BV$ returns $1$ for each ground atom, the set $\Act = \{a_1,\ldots,a_m\}$ is set s.t. each action in $\Act$ corresponds to an element in $\calh$ and each $a_i$ is defined as follows: $a_i(p) = \bigcup_{x_j \in H_i}\{g_j(p) \}$.  The cost function $\CM$ returns $1$ for each action-point pair, and finally, we set $s_0 = \emptyset$, $IC = \emptyset$, $k = K$, $\bc = K$.$\hfill\Box$\\


One may think that one can solve GOPs efficiently in practice by using fully polynomial time approximation schemes (FPTAS).  However, by the nature of our constructions used in the NP-hardness results, this is not possible for either type of GOP under accepted theoretical assumptions.

\begin{theorem}
\label{gbgop-non-apprx}
If for some $\epsilon >0$, there is a PTIME algorithm to approximate GBGOP within $(1-\epsilon) \cdot \ln(|\APPall|)$, then 
$NP \subset TIME(|\APPall|^{O(\lg\lg |\APPall|)})$ (NP has a slightly super-polynomial algorithm).
\end{theorem}
\noindent Follows from Theorem~\ref{gbgop-nph} and  \cite[Theorem 4.4]{feige98}.$\hfill\Box$\\

\begin{theorem}
\label{gop-no-apprx}
Finding an optimal solution to BMGOP cannot be approximated in PTIME within a ratio of $\frac{e-1}{e}+\epsilon$ (approx. $0.63$) for some $\epsilon >0$  (where $e$ is the inverse of the natural log) unless \textbf{P=NP}, even when $IC = \emptyset$.
\end{theorem}
\noindent Follows from Theorem~\ref{gop-nph} and \cite[Theorem 5.3]{feige98}.$\hfill\Box$\\

Next, under some reasonable assumptions, the decision problems for GBGOP/BMGOP are in-NP.

\begin{theorem}
\label{gbgop-npc}
Given GBGOP $(\calm,s_0,\Act,\CM,IC,\bc,\inset,$\\$\outset)$, if the cost function and all actions $a \in \Act$ can be polynomially computed, then determining if there is a solution $SOL$ for the instance of the GBGOP s.t. for some real number $k$, $|SOL| \leq k$ is in-NP.
\end{theorem}

\begin{theorem}
\label{gop-npc}
Given BMGOP $(\calm,s_0,\BV,\Act,\CM,IC,k,\bc)$, if the cost function, benefit function, and all actions $a \in \Act$ can be polynomially computed, then determining if there is a solution $SOL$ for the instance of the BMGOP s.t. for some real number $val$, $\sum_{A_i \in appl(SOL,s_0)}b_i \geq val$ is in-NP.
\end{theorem}

As stated earlier, a GBGOP may also be viewed as an abductive inference problem.  Even though finding a solution (not necessarily optimal) to a GBGOP can trivially be conducted in PTIME\footnote{Return the set $\{(a_i,p_i) \in \APPall | a_i(p_i) \cap \outset = \emptyset \}$}, counting the number of solutions is $\#$P-complete.  This counting problem is difficult to approximate.

\begin{theorem}
\label{shp-compl}
Counting the number of solutions to a GBGOP (under the assumptions of Theorem~\ref{gbgop-npc}) is $\#$P-complete.
\end{theorem}
\noindent\textit{Proof Sketch.} The MONSAT problem~\cite{roth96} takes a set $C$ of $m$ clauses of $K$ disjunct ed literals (no negation) over set $L$ of atoms (size $n$) and outputs ``yes'' iff there is a subset of $L$ that satisfies all clauses in $C$.  We encode this into finding a GBGOP as follows: $\calg = \{g_1,\ldots,g_m\}$ - each predicate in $\calg$ corresponds to an clause in $C$ (predicate $g_j$ corresponds with clause $\phi_j$), $\calm$ consists of a single point, $p$, $\Act = \{a_1,\ldots,a_n\}$ - each action in $\Act$ corresponds to an element in $L$ (action $a_i$ corresponds with literal $\ell_i$).  Each $a_i$ is defined as follows: $a_i(p) = \{ g_j(p) | \{\ell_i\} \models \phi_j\}$, $\CM$ returns $1$ for all action-point pairs, $s_0 = \emptyset$, $IC = \emptyset$, $\bc = n$, $\inset = \bigcup_{g_i \in \calg}\{ g_i(p)\}$, $\outset = \emptyset$.  Based on this PTIME reduction we show a 1-1 correspondence to MONSAT.  Hence, we can parsimoniously reduce the counting version of MONSAT (number of solutions) to the counting version of GBGOP (number of solutions).  As the counting version of MONSAT is $\#$P-hard by \cite{roth96}, we have shown that $\#$P-hardness of the counting version of GBGOP.  As there is an obvious bound on the number of solutions to a GBGOP, and as the solutions are verifiable in PTIME, membership in $\#$P follows.$\hfill\Box$\\

\begin{theorem}
\label{cnt-apprx}
For $\epsilon >0$, approximating the number of solutions to a GBGOP within a factor of $2^{|\APPall|^{1-e}}$ is NP-hard.
\end{theorem}
\noindent Follows from Theorem~\ref{shp-compl} and Theorem 3.2 of \cite{roth96}.$\hfill\Box$\\

Due to this issue with achieving a good approximation of the counting version, in this paper we shall focus only on determining a single optimal solution to a GBGOP - rather than all solutions.
\section{Integer Programs for Solving GOPs}
\label{algo:sec}
In this section, we present an integer  programming (IP) algorithms for both GBGOP and BMGOP which provide exact solutions.  Given a GBGOP, the IP associates an integer-valued variable $X_i$ with each action-point pair $(a_i,p_i)\in \APPall$ where $a_i(p_i) \cap \outset = \emptyset$. Intuitively,  $X_i=1$ denotes that action $a_i$ is performed at point $p_i$.

\begin{definition}[GBGOP-IP]
Let set $R = \{(a_i,p_i) \in \APPall | a_i(p_i) \cap \outset = \emptyset \}$.  For each action-point pair $(a_i,p_i) \in R$, create variable $X_{i} \in \{0,1\}$.
\begin{small}
\begin{eqnarray}
\min \, \sum_{i=1}^{|R|}X_{i} &\\
\label{gbc1}\textit{s.t.} \, \mathop{\sum}_{a_j(p_j) | A_i \in a_j(p_j)}X_{j} \geq 1 & \forall A_i \in \inset-s_0\\
\label{gbc2} \, \mathop{\sum}_{(a_i,p_i) \in R}c_{i}\cdot X_{i} \leq \bc&\\
\label{gbc3} \, \mathop{\sum}_{(a_i,p_i) \in \APP}X_i \leq 1 & \forall (\APP \hookleftarrow \chi) \in IC_{s_0}
\end{eqnarray}
\end{small}
\end{definition}

The objective function minimizes the total number of action-point pairs.  Constraint (\ref{gbc1}) ensures that every ground atom in $\inset$ (that does not appear in the initial state) is caused by at least one of the selected action-point pairs. Constraint (\ref{gbc2}) enforces the constraint on cost. Constraint (\ref{gbc3}) ensures that the integrity constraints are satisfied.  Next we present our integer constraints for a BMGOP where the IP associates an integer-valued variable $X_i$ with each action-point pair $(a_i,p_i)\in \APPall$, and an integer-valued variable $Y_j$ with each ground atom $A_j \in B_\call-s_0$. The intuition for the $X_i$ variables is the same as in GBGOP-IP.

\begin{definition}[BMGOP-IP]
For each action-point pair $(a_i,p_i) \in \APPall$, create variable $X_{i} \in \{0,1\}$.  For each $A_i \in B_\call-s_0$ create variable $Y_i \in \{0,1\}$.
\begin{small}
\begin{eqnarray}
\label{c0} \max \, \sum_{A_i \in s_0}b_{i}+\sum_{i=1}^{|B_\call|-|s_0|}b_{i}\cdot Y_{i} &\\
\label{c1}\textit{s.t.} \, \mathop{\sum}_{a_j(p_j) | A_i \in a_j(p_j)}X_{j} \geq Y_{i} & \forall A_i \in B_\call-s_0\\
\label{c2} \, \mathop{\sum}_{(a_i,p_i) \in \APPall}X_{i} \leq k &\\
\label{c3} \, \mathop{\sum}_{(a_i,p_i) \in \APPall}c_{i}\cdot X_{i} \leq \bc&\\
\label{c4} \, \mathop{\sum}_{(a_i,p_i) \in \APP}X_i \leq 1 & \forall (\APP \hookleftarrow \chi) \in IC_{s_o}
\end{eqnarray}
\end{small}
\end{definition}
In the above IP, the objective function looks at each ground atom and sums the associated benefit if the associated $Y_i$ variable is $1$ - meaning that atom $A_i$ is true after the actions are applied.  Constraint (\ref{c1}) effectively sets a $Y_i$ variable to $1$ if an action that causes $A_i$ to be true occurs.  Constraint (\ref{c2}) enforces the cardinality requirement.  Constraints~\ref{c3}-\ref{c4} mirror constraints~\ref{gbc2}-\ref{gbc3} of GBGOP-IP.  The result below shows that a solution $\sigma$ to the above IPs\footnote{A solution to GBGOP-IP or BMGOP-IP is an assignment of values to variables that optimizes the objective function. Thus, a solution can be described as a set of equations assigning values to the variables $X_i,Y_j$.}, when restricted to the $X_i$ variables, provides an immediate solution to the GOP.

\begin{proposition}
\label{gbgop-corr-prop}
Suppose $\Gamma$ is a GBGOP (resp. BMGOP) and $IP(\Gamma)$ is its corresponding integer program (GBGOP-IP, resp. BMGOP-IP). Then:
\begin{enumerate}
\item If $SOL$ is  a solution to $\Gamma$, then there is a solution $\sigma$ of $IP(\Gamma)$ such that $\sigma\supseteq
\{ X_i=1\: |\: (a_i,p_i)\in SOL\}$.
\item If $\sigma$ is a solution to $IP(\Gamma)$, then there is a solution $SOL$ to $\Gamma$ such that $\{ X_i=1\: |\: (a_i,p_i)\in SOL\}\subseteq \sigma$.
\end{enumerate}
\end{proposition}

As integer programming is NP-complete, any algorithm to solve a GOP using GBGOP-IP or BMGOP-IP using an IP solver will take exponential time.  We note that for GBGOP-IP, the number of variables is fairly large -- $O(|\{(a_i,p_i) \in \APPall | a_i(p_i) \cap \outset = \emptyset \}|)$ variables and $O(|\inset-s_0|+|IC_{s_0}|+1)$ constraints. BMGOP-IP has even more variables - (though not exponential) - $O(|\calm|\cdot(|\Act|+|\calg|))$ variables 
and  $O(|\calm|\cdot |\calg|+|IC_{s_0}|+2)$ constraints.  
However, BMGOP-IP has only packing constraints.\footnote{It is trivial to eliminate constraint~\ref{c1} and re-write \ref{c0} as a non-linear objective function.}
We also note the GBGOP-IP has both covering ($\geq$) and packing ($\leq$) constraints - another source of complexity. 

\section{Correct Variable Reduction for GBGOP-IP}
\label{var-red-sec}

The set of integer constraints for GBGOP has $O(|R|)$ variables where $R\subseteq \APPall$.  We show how to correctly reduce the number of variables by considering only a subset of $R$ - thereby providing a smaller integer program.  Our intuition is that an optimal solution $SOL$ is an \textit{irredundant} cover of $\inset$ meaning there is no subset $SOL' \subset SOL$ that is also a solution.  Hence, we can discard certain elements of $R$ that cannot possibly be in an optimal solution.  First, for a given GBGOP $\Gamma=(\calm,s_0,\Act,\CM,IC,\bc,\inset,\outset)$, we introduce  $\Qset  =  \{ \APP | (\APP \hookleftarrow \chi)\in IC_{s_0} \wedge (a,p)\in \APP \}$ and the set of ground atoms each action-point pair affects  $\Affset  = a_i(p_i) \cap (\inset-(\inset \cap s_0))$.  We can now define a \textit{reduced action-point set}.

\begin{definition}[Reduced Action-Point Set]
\label{red-set}
Given GBGOP $\Gamma=(\calm,s_0,\Act,\CM,IC,\bc,\inset,\outset)$ and set $R = \{(a_i,p_i) \in \APPall | a_i(p_i) \cap \outset = \emptyset \}$, we define \textbf{reduced action-point set} $R^{*} =  \{ (a_i,p_i) \in R | \not\exists (a_j,p_j)\in R \textit{ s.t. }$\\$(c_j \leq c_i) \wedge(\Qsetj\subseteq \Qseti) \wedge (\Affseti \subseteq \Affsetj)\}$
\end{definition}

\begin{example}
Consider the campaign scenario last discussed in Example~\ref{ic-ex}.  Suppose the candidate wants to optimize the following GBGOP: $\Gamma=(\exm,\exs,\exA,\excs,\exic,4,\insetex,\emptyset)$ where each $A \in \insetex$ has the form $exposure(p)$ where $p$ is a point in one of the two dashed rectangles in Figure~\ref{fig1}.  Note that as map $\exm$ contains $187$ points, $|\Act|=3$, and $\outset=\emptyset$, the cardinality of $R$ is $561$.  By contrast, the set $R^{*}$ consists of only $7$ elements, $1.2 \%$ of the size of $R$.  Here $R^{*}=\{ (nor,(5,4)), (nor,(5,3)), (nor,(5,2)), (nor,(10,8)),$\\$(nor,(10,7)), (nor,(10.6)), (appeal_1,(4,3))\}$
\end{example}

Intuitively, all elements in $R^{*}$ are preferable for membership in an optimal solution over $R-R^{*}$ as they cost less, result in the same changes to the state, and occur in the same or fewer integrity constraints.  Set $R^{*}$ can be found in quadratic time with a naive algorithm - an operation that is likely dominated by solving or approximating GBGOP-IP.  The next lemma says that $R^{*}$ must contain an optimal solution.any optimal solution to a GBGOP.  This can then be used to correctly reduce the number of variables in GBGOP-IP.

\begin{lemma}
\label{subset-lem}
Given GBGOP $\Gamma=(\calm,s_0,\Act,\CM,IC,\bc,\inset,\outset)$, for any optimal solution $SOL \subseteq R$, there is an optimal solution $SOL' \subseteq R^{*}$.\footnote{\noindent\textit{Proof Sketch.} We show this by proving that for any set $W = SOL \cap (R - R^{*})$, there is some set $W'\subseteq R^{*}-(R^{*} \cap SOL)$ s.t. $(SOL-W) \cup W'$ is also a solution.}
\end{lemma}

\begin{proposition}
\label{red-var-prop}
Suppose $\Gamma$ is a GBGOP and $IP(\Gamma)$ is its corresponding integer program.  We can create such a program with a variable for every element of $R^{*}$ (instead of $R$) and Proposition~\ref{gbgop-corr-prop} still holds true.
\end{proposition}

\section{The \textsf{BMGOP-Compute} Algorithm}
\label{mu-alg-sec}
While BMGOP-IP can solve a BMGOP exactly, doing so is computationally intractable.  We now present an approximation algorithm that runs in PTIME but provides a lower approximation ratio than proved in Theorem~\ref{gop-no-apprx}.  First, we show that a BMGOP reduces to an instance of submodular maximization problem\footnote{Suppose $Z$ is a set. A function $f:2^Z\rightarrow \mathbb{R}$  is said to be \emph{submodular} iff for all $Z_1,Z_2$ such that $Z_1\subseteq Z_2$ and all $z\notin Z_2$, it is the case that $f(Z_1\,\cup\,\{z\})-f(Z_1)\geq f(Z_2\,\cup\{z\})-f(Z_2)$, i.e. the incremental value of adding $z$ to the smaller set $Z_1$ exceeds the incremental value of adding it to the larger set $Z_2$. Here, $\mathcal{R}$ denotes the reals.} with respect to packing constraints.  We then leverage some known methods~\cite{gamzu10} to solve such problems and develop a fast, deterministic algorithm to approximate BMGOP with an approximation bounds.  
Given BMGOP $\Gamma=(\calm,s_0,\BV,\Act,\CM,IC,k,\bc)$, consider the objective function in BMGOP-IP.  We can write that function as a mapping from action-point pairs to reals.  We denote this function (specific for BMGOP $\Gamma$) as $f_\Gamma : 2^{\APPall} \rightarrow \Re^{+}$, where $f_\Gamma(S) = \sum_{A_i \in \textsf{appl}(S,s_0)}b_i
$, which has certain properties.
\begin{small}
\begin{equation}
\label{app-fcn}
f_\Gamma(S) = \sum_{A_i \in \textsf{appl}(S,s_0)}b_i
\end{equation}
\end{small}

We now show that this function $f_\Gamma$ is submodular and has some other nice properties as well.

\begin{proposition}
For BMGOP $\Gamma$, function $f_\Gamma$ is:
\textsf{(i)} submodular, \textsf{(ii)} monotonic, i.e. $Z_1\subseteq Z_2\rightarrow f_\Gamma(Z_1)\leq f_\Gamma(Z_2)$ and \textsf{(iii)} under the condition $\forall A_i \in B_\call$, $b_i=0$, we have $f_\Gamma(\emptyset)=0$.\footnote{Henceforth, we will assume this condition to be true.}
\end{proposition}
\noindent\textit{Proof Sketch.} Consider $S \subseteq S' \subseteq \APPall$ and $(a,p) \notin S'$.  We must show $f_\Gamma(S \cup \{(a,p)\}) - f_\Gamma(S) \geq f_\Gamma(S' \cup \{(a,p)\}) - f_\Gamma(S')$.  Suppose, BWOC $f_\Gamma(S \cup \{(a,p)\}) - f_\Gamma(S) < f_\Gamma(S' \cup \{(a,p)\}) - f_\Gamma(S')$.  Then, by Equation~\ref{app-fcn}, we have $\sum_{A_i \in appl(S \cup \{(a,p)\},s_0)-appl(S,s_0)}b_i < \sum_{A_i \in appl(S' \cup \{(a,p)\},s_0)-appl(S',s_0)}b_i$.  However, by the definition of $appl$, we have $appl(S \cup \{(a,p)\},s_0)-appl(S,s_0) \supseteq appl(S' \cup \{(a,p)\},s_0)-appl(S',s_0)$, which is a contradiction.$\hfill\Box$\\

As our objective function is submodular, and constraints~\ref{c2}-\ref{c4} are linear packing constraints, any instance of a BMGOP can be viewed as maximization of a submodular function wrt linear packing constraints and hence, methods to solve such problems can be used here. The \textsf{BMGOP-Compute} algorithm leverages this idea and illustrated in Example~\ref{gmu-ex}.

\begin{algorithm}\textsf{BMGOP-Compute}\\
\noindent INPUT: BMGOP $(\calm,s_0,\BV,\Act,\CM,IC,k,\bc)$\\
\noindent OUTPUT: $SOL \subseteq \APPall$
\begin{small}
\begin{enumerate}
\item Set $SOL = \emptyset$, $\delta$ to be an infinitesimal, \\ and set $\lambda = e^{2-\delta}\cdot (2+|IC_{s_0}|)$.
\item Set $w' = 1/k$ and $w'' = 1/\bc$.  For each $(\APP_i \hookleftarrow \chi_i) \in IC_{s_0}$, set $w_i = 1/(2-\delta)$.
\item\label{big-lp} While $k\cdot w' + \bc\cdot w'' + (2-\delta)\cdot\sum_i w_i \leq \lambda$ and $SOL \neq \APPall$
	\begin{enumerate}
	\item\label{qty} Let $(a_j,p_j) \in \APPall-SOL$ have minimal\\
	$\frac{w'+w''\cdot c_j+\sum_{i | (a_j,p_j)\in \APP_i}w_i}{(\sum_{A_i \in appl(SOL\cup \{(a_j,p_j)\},s_0)}b_i)-(\sum_{A_i \in appl(SOL,s_0)}b_i)}$
	\item $SOL = SOL \cup \{(a_j,p_j)\}$
	\item Set $w' = w' \cdot \lambda^{1/k}$, $w'' = w'' \cdot \lambda^{c_j/\bc}$ and for each integrity constraint $i$ s.t. $(a_j,p_j) \in \APP_i$, set\\ $w_i = w_i \cdot \lambda^{1/(2-\delta)}$
	\end{enumerate}
\item\label{chk-ln} If $SOL$ is not a valid solution then
	\begin{enumerate}
	\item If 	$\sum_{A_i \in appl(SOL - \{(a_j,p_j)\},s_0)}b_i \geq$\\
	$\sum_{A_i \in appl(\{(a_j,p_j)\},s_0)}b_i$,\\
	then $SOL = SOL - \{(a_j,p_j)\}$
	\item Else $SOL = \{(a_j,p_j)\}$
	\end{enumerate}
\item Return $SOL$
\end{enumerate}
\end{small}
\end{algorithm}

\begin{example}
\label{gmu-ex}
Following Example~\ref{ic-ex}.  Suppose the candidate wants to optimize BMGOP: $(\exm,\exs,\exb,\exA,\excs,\exic,3,2)$.  In this case, we will set $\delta=0.001$.  He wishes to find a set of $3$ action-point pairs to optimize his exposure.  \textsf{BMGOP-Compute} sets $\lambda=22.14$, $w'=0.33$, $w''=0.50$, and $w_1 = 0.50$ in lines 1 and 2.  In the first iteration of the loop at line~\ref{big-lp}, it finds the action-point pair that minimizes the quantity at line~\ref{big-lp} is $(appeal_1,(4,3))$ - which has the associated value $0.073$.  Note, other action-point pairs with low values are  $(appeal_2,(10,7))$ with $0.083$ and $(nor,(15,6))$ also with $0.083$.  It then adds $(appeal_1,(4,3))$ to $SOL$ and updates $w'=0.93$, $w''=1.09$, and $w_1 = 2.35$.  On the next iteration, the \textsf{BMGOP-Compute} picks $(nor,(15,6))$, which now has a value of $0.164$.  During this iteration, the value of $(appeal_2,(10,7))$ has increased substantially - to $0.294$, so it is not selected.  At the end of the iteration, $w'$ is updated to $2.611$ and $w''$ is updated to $2.364$.  As $(nor,(15,6))$ does not impact the lone integrity constraint, the value $w_1$ remains at $2.354$.  In the third iteration, \textsf{BMGOP-Compute} selects $(nor,(15,9))$ which has a value of $0.421$.  Again, the value of $(appeal_2,(10,7))$ has increased - but this time only to $0.472$.  \textsf{BMGOP-Compute} re-calculates $w'=7.331$, $w''=5.128$ and $w_1$ remains at $2.354$.  On the last iteration, \textsf{BMGOP-Compute} picks $(appeal_2,(10,7))$ as it has the lowest value --  $0.942$.  After this fourth iteration, it updates $w'=20.589$, $w''=11.124$, and $w_1=11.0861$ - which now total to $42.799$ -- exceeding $\lambda$ ($22.14$) -- causing \textsf{BMGOP-Compute} to exit the outer loop.  Now $SOL$ has $4$ elements, exceeding the cardinality constraint (as well as the integrity constraint).  The checks done in line~\ref{chk-ln} remove $(appeal_2,(10,7))$ from $SOL$ - making the result feasible.  \textsf{BMGOP-Compute} returns $\{(appeal_1,(4,3)), (nor,(15,6)_, (nor,(15,9))\}$ which causes the benefit to be $45$.
\end{example}

\begin{proposition}
Suppose $\Gamma$ is a BMGOP and $SOL$ is the set returned by \textsf{BMGOP-Compute}. Then $SOL$ is a solution to $\Gamma$.\footnote{Here, $SOL$ is not necessarily an optimal solution.}
\end{proposition}

Next, we sho \textsf{BMGOP-Compute} runs in PTIME.

\begin{proposition}
\label{BMGOP-mu-cmplx}
\textsf{BMGOP-Compute} runs in $O(k \cdot |\calm| \cdot |\Act|\cdot |IC_{s_0}|)$ time.
\end{proposition}
\noindent\textit{Proof Sketch.} Clearly, the outer loop can iterate no more than $k$ times.  The inner loop iterates for each element of $\APPall$ - hence requiring time $O(|\calm| \cdot |\Act|)$.  There are some additional operations that require $O(|IC_{s_0}|)$ time, however, they are dominated under the assumption that $|\calm| \cdot |\Act| >> |IC_{s_0}|$, which we expect in our application.$\hfill\Box$\\

\begin{figure}
	\centering
		\includegraphics[scale=0.25]{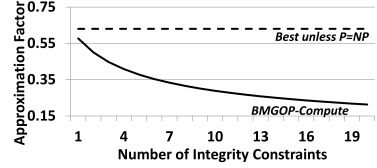}		
	\caption{$|IC_{s_0}|$ vs. approximation ratio.}
	\label{apprx-fig}
\end{figure}

The following important theorem states that \textsf{BMGOP-Compute} provides an approximation guarantee. Because of Theorem~\ref{gop-no-apprx} and as \textsf{BMGOP-Compute} is polynomial, 
we know that this approximation guarantee cannot be as good as $\frac{e-1}{e}+\epsilon$. The result leverages Theorem 1.1 of \cite{gamzu10} together with the above theorems.  By this result, the approximation factor of \textsf{BMGOP-Compute} depends on $|IC_{s_0}|$.  We illustrate this relationship, in Figure~\ref{apprx-fig}.  For our target applications, we envision $|IC_{s_0}|\leq 20$.

\begin{theorem}
\label{apprx-thm}
Under the assumption that $k,\bc \geq 2-\delta$, \textsf{BMGOP-Compute} provides a solution within a factor of $\frac{1}{(2+|IC_{s_0}|)^{1/(2-\delta)}}$ (where $\delta$ is an infinitesimal) of optimal.
\end{theorem}
\noindent\textit{Proof Sketch.}  \textsf{BMGOP-Compute} follows from Algorithm 1 of \cite{gamzu10} which optimizes a submodular function subject to $m$ packing constraints within $\frac{1}{m^{1/W}}$ where $W$ is the minimum width of the packing constraints - defined as the minimum of the size of the constraint divided by the cost of an element.  For constraint~\ref{c2}, the $W=k$.  For constraint~\ref{c3}, the $W\geq\bc$.  We can replace constraint~\ref{c4} with: $\mathop{\sum}_{(a_i,p_i) \in \APP_j}X_i \leq 2-\delta$   $\forall (\APP_j \hookleftarrow \chi_j) \in IC_{s_o}$ which maintains correctness as two variables to set to $1$ and exceeds $2-\delta$.  The new constraint has width $2-\delta$, which, is the minimum.  We then apply Theorem 1.1 of \cite{gamzu10}.$\hfill\Box$\\


\noindent\textbf{Discussion.} We note that while a BMGOP reduces to the maximization of a submodular or linear function wrt linear packing constraints, there are other algorithms available besides the multiplicative update algorithm of \cite{gamzu10}.  However, we feel that this is likely the best approach for several reasons that we list below.
\begin{enumerate}
\item{The approximation ratio achieved by the multiplicative-update algorithm matches the best approximation ratio achievable for maximizing a linear function wrt linear packing constraints (see \cite{srin95}), hence, it is unlikely that a better approximation ratio can be achieved using such a technique.}
\item{Other methods (such as those presented in \cite{srin95}) require solving a relaxation of the associated MILP.  In our case, such an operation would take $O((|\calm|\cdot(|\Act|+|\calg|))^{3.5})$ time (as a consequence of the number of variables in BMGOP-MILP and the results of \cite{karm1984}).  This is significantly more expensive than the  $O(k \cdot |\calm| \cdot |\Act|)$ of \textsf{BMGOP-MU} (see Proposition~\ref{BMGOP-mu-cmplx}).  If the map, $\calm$ is very large, solving a relaxation of BMGOP-MILP may be unrealistic on most hardware.}
\item{The algorithm \textsf{BMGOP-MU} is totally deterministic, which allow us to avoid the issue de-randomization.}
\item{The algorithm \textsf{BMGOP-MU} is guaranteed to provide a solution that meets constraints~\ref{c2}-\ref{c4} - as opposed to only meeting them probabilistically.}
\end{enumerate}
\section{Related Work and Conclusions}

Though spatial reasoning has been studied extensively in AI \cite{ch01,egs91,rn99,ly03}, many of the paradigms that have emerged for such reasoning are  \emph{qualitative} in nature. Such qualitative spatial reasoning efforts include the influential region connection calculus for qualitative reasoning about space.  There has also been work on quantitative methods for reasoning about space \cite{rpps10}  which contains articles on spatial reasoning in the presence of uncertainty using both logical and fuzzy methods. Spatial reasoning with quantitative information has been studied extensively in image processing \cite{wa96,sr95}.

However, unlike this vast body of work, this paper focuses on a different problem. Suppose we are dealing with a map $\calm$, a cost function $\CM$, a set $\Act$ of possible actions, a bound on the cost $\bc$, and a bound on the number of actions we can take, what set of actions should be taken so as to optimize a given objective function. Two versions of this problem are studied in this paper - GBGOP and BMGOP which differ in what they optimize. Both problems are proved to be NP-hard (NP-complete under realistic assumptions) and we further prove that the number of solutions to GBGOP is \#P-complete. We also find limits on approximating an optimal solution to BMGOP and GBGOP (in PTIME) under accepted theoretical assumptions.  We develop integer programming formulations of both problems and then present a way of simplifying the IP for GBGOP. We further present the \textsf{BMGOP-Compute} algorithm for BMGOP and show that it is polynomial and has a guaranteed approximation ratio (though not high enough to contract the NP-hardness result).

\section{Conclusion}

In this paper, we introduced ``geopspatial optimization problems'' or GOPs that aide the user in taking certain actions over a geographic region.  We showed these problems to be NP-hard and provided integer constraints.  For the goal-based variant, we correctly reduce the number of variables.  For the benefit-maximizing variant, we provide an approximation algorithm.  In future work, we look to implement this framework and explore methods to achieve further scalability, as well as utilize geo-located social network data to establish relationships among locations in order to better implement action-point pairs and integrity constraints.




\end{document}